\documentclass[preprint,prd,
amsmath,amssymb,amsthm,nofootinbib,superscriptaddress,longbibliography]{revtex4-2}

\usepackage{graphicx}   
\usepackage[
colorlinks=true,        
citecolor=blue,         
linkcolor=blue,         
urlcolor=blue ,          
]{hyperref}  
\usepackage{tabulary}
\usepackage{color}      
\usepackage{orcidlink}
\usepackage{multirow}
\usepackage{footnote}
\usepackage{floatrow}
\usepackage{float} 
\usepackage{bm}

\newcommand{\m}{\mathrm}

\begin{document}
\title{Photon rest mass from localized fast radio bursts with  improved distribution of dispersion measure from extragalactic gas} 

\author{Yuchen Zhang\orcidlink{0009-0009-3644-2082}}
\affiliation{Department of Physics, Hunan Research Center of the Basic Discipline for Quantum Effects and Quantum Technologies, and Key Laboratory of Low-Dimensional Quantum Structures and Quantum Control of Ministry of Education,
Hunan Normal University, Changsha, Hunan 410081, China}

\author{Yang Liu\orcidlink{0000-0003-2721-2559}}
\affiliation{Purple Mountain Observatory, Chinese Academy of Sciences, No. 10 Yuanhua Road, Nanjing 210023, China}

\author{Hongwei Yu\orcidlink{0000-0002-3303-9724}}
\email{hwyu@hunnu.edu.cn}
\affiliation{Department of Physics, Hunan Research Center of the Basic Discipline for Quantum Effects and Quantum Technologies, and Key Laboratory of Low-Dimensional Quantum Structures and Quantum Control of Ministry of Education,
Hunan Normal University, Changsha, Hunan 410081, China}

\author{Puxun Wu\orcidlink{0000-0002-9188-7393}}
\email{pxwu@hunnu.edu.cn}
\affiliation{Department of Physics, Hunan Research Center of the Basic Discipline for Quantum Effects and Quantum Technologies, and Key Laboratory of Low-Dimensional Quantum Structures and Quantum Control of Ministry of Education,
Hunan Normal University, Changsha, Hunan 410081, China}

\begin{abstract}
The assumption that photons are massless is a foundational postulate of modern physics, yet it remains subject to experimental verification. Fast radio bursts (FRBs), with their cosmological distances and precisely measured dispersion, offer an excellent laboratory for testing this hypothesis. In this work, we propose an improved distribution function for the dispersion measure arising from extragalactic gas and demonstrate that it provides an excellent fit to mock data. We then apply this distribution to constrain the photon rest mass under the $\Lambda$CDM, $w$CDM, and $w_{0}w_{a}$CDM cosmological models, the last of which is favored by recent DESI baryon acoustic oscillation observations. The corresponding $1\sigma$ upper limits on the photon mass are found to be $4.83\times10^{-51}\,\mathrm{kg}$, $4.71\times10^{-51}\,\mathrm{kg}$, and $4.86\times10^{-51}\,\mathrm{kg}$, respectively, which are the most stringent constraints derived from FRBs to date. These results   demonstrate that FRBs provide robust and reliable constraints, and offer strong empirical support for the massless nature of the photon.
\end{abstract}

\maketitle

\section{Introduction} 

A central postulate  of Einstein’s special relativity is the invariance of the speed of electromagnetic waves~\cite{Einstein:1905vqn}: the speed of light in  vacuum is constant,  independent of the motion of either the source or the observer.   This principle implies that photons must have zero rest mass since a non-zero photon mass would result in  a frequency-dependent  group velocity of photons,  causing higher-frequency components to travel faster than the lower-frequency ones. Consequently, photons emitted simultaneously from the same source will exhibit a frequency-dependent time delay upon arrival, providing  an opportunity to experimentally test whether photons are massless.

Numerous ground tests have been conducted to measure variations in the speed of light in order to examine the photon rest mass~\cite{Williams:1971ms, Kroll:1971wi, Chernikov:1992sb, Lakes:1998mi, Luo:2003rz}. In recent years, astrophysical observations, especially gamma-ray bursts, have emerged as powerful probes for  the relative propagation speeds of electromagnetic signals over cosmological distances~\cite{Schaefer:1998zg, Wei:2016jgc, Bartlett:2021olb, lovell1964relative, warner1969wavelength, zhang2016constraining, wei2018robust}.  Because light from distant sources travels for billions of years, even minuscule variations in photon speed can accumulate into measurable time delays, enabling exceptionally stringent constraints on a possible photon mass.

Fast radio bursts (FRBs), enigmatic radio transients discovered in recent decades~\cite{Lorimer:2007qn}, are generally accepted to have a cosmic origin. Given that residual baryons are believed to exist in diffused state in the intergalactic medium (IGM), the ionized electrons arising from these baryons interact with the radio signals, leading to a time delay in their arrival. This delay is quantified by the dispersion measure (DM). 
There have already been numerous applications of the dispersion measures (DMs) of localized fast radio bursts (FRBs) to  explore the cosmic reionization history~\cite{Zheng:2014rpa, Caleb:2019apf, Beniamini:2020ane, Hashimoto:2021tyk, Wei:2024tpv, Shaw:2024slf} and constrain cosmological parameters,  such as the Hubble constant $H_{0}$~\cite{Li:2017mek, hagstotz2022new, james2022measurement, Liu:2022bmn, Fortunato:2024hfm, Piratova-Moreno:2025cpc, Wu:2021jyk}, the present cosmic baryon density parameter $\Omega_{b0}$ and the baryon fraction in the diffused state (i.e.~$f_\m{d}$)~\cite{Connor:2024mjg,Lemos:2025bgy, Li:2020qei, Wang:2022ami, Deng:2013aga, Ravi:2018ose, Munoz:2018mll, Li:2019klc, Walters:2019cie, Wei:2019uhh, Macquart:2020lln, Yang:2022ftm, Lin:2023opv, Liu:2025fdf, Zhang:2025yhi},  and so on. 
When we assume a non-zero rest mass for the photon, the DM will include an additional contribution from the mass term, thereby creating a promising tool to constrain the photon rest mass using FRBs~\cite{Bentum:2016ekl, Lemos:2025qyh, Chang:2024hnn, Ran:2024avn, Wang:2024rgu, Shao:2017tuu, Wu:2016brq, Wei:2020wtf, Bonetti:2016cpo, Bonetti:2017pym,  Xing:2019geq, Chang:2022qct}.

A key ingredient in FRB cosmology is the distribution function of the extragalactic DM,  $\m{DM_{cosmic}}$,  which accounts for contributions from the IGM and intervening halos. A widely used $\m{DM_{cosmic}}$ distribution  was proposed based on semi-analytic models and hydrodynamic simulations~\cite{Macquart:2020lln, Zhang:2020xoc}, with fixed parameters $\alpha=\beta=3$. Recently, Konietzka et al.~\cite{Konietzka:2025kdr} introduced  a new method for measuring $\m{DM_{cosmic}}$ in the IllustirsTNG cosmic simulation. They  found that  previous work~\cite{Zhang:2020xoc} had misestimated the standard deviation and higher moments of the $\m{DM_{cosmic}}$ distribution by over 50\%. They constructed an analogue distribution of $ \m{DM_{cosmic}}$ similar to that in~\cite{Macquart:2020lln}  and  confirmed that 	this distribution,   with parameters $\alpha\approx 1$ and $\beta\approx 3.3$,  provides good fits across all redshifts. However, the distribution function presented in~\cite{Konietzka:2025kdr}  does not explicitly contain information of our universe, and  therefore cannot be used directly to estimate cosmological parameters. 

In this paper, we first construct an improved distribution function of $\m{DM_{cosmic}}$ that incorporates  cosmological information, allowing the parameters $\alpha$ and $\beta$ to vary with redshift. We find that our function is more consistent with mock data compared to those proposed in~\cite{Macquart:2020lln, Konietzka:2025kdr}. 
		
We then utilize our distribution function to constrain the photon rest mass using FRB observations. Since FRBs are predominantly emitted from extragalactic sources, it is essential to select a specific cosmological model. The $\Lambda$CDM model, which consists of cosmological constant  $\Lambda$ and cold dark matter, is typically employed in such analyses~\cite{Wu:2016brq, Bonetti:2016cpo, Bonetti:2017pym, Shao:2017tuu, Xing:2019geq, Wei:2020wtf, Wang:2021nrl, Chang:2022qct, Lin:2023jaq, Wang:2023fnn}. However, recent  baryon acoustic oscillation (BAO) measurements from the Dark Energy Spectroscopic Instrument (DESI) Data Release 2 (DR2) indicate a strong  preference for dynamical dark energy over the cosmological constant dark energy with a confidence level (CL) exceeding 3$\sigma$~\cite{DESI:2025zgx}. This  motivates us to reanalyze photon mass constraints within cosmological models accommodating dynamical dark energy, such as the $w$CDM and $w_{0}w_{a}$CDM models~\cite{Chevallier:2000qy, Linder:2002et}.
To constrain the parameters of these cosmological models, we combine data from  Pantheon+ type Ia supernovae (SN Ia)~\cite{Scolnic:2021amr}, BAO from DESI~\cite{DESI:2025zgx}, and the cosmic microwave background (CMB) from Planck~\cite{Planck:2018vyg} .

This paper is organized as follows. In Sec.~\ref{secmeth}, we provide a detailed description of the effective DM arising from a non-zero photon mass. In Sec.~\ref{secdata}, we present  the observational datasets used, including FRB, CMB, BAO, and SN data. Our cosmological constraints are reported in Sec.~\ref{result}. Finally, we summarize our conclusions and discussions  in Sec.~\ref{seccon}.

\section{DM from photon mass}
\label{secmeth} 
	
If photons possess a non-zero rest mass $m_\gamma$, the energy of a single photon is
\begin{eqnarray}\label{eq1}
	E = \sqrt{p^2 c^2 + m_{\gamma}^2 c^4},
\end{eqnarray}
where $p$ is the photon momentum and $c$ is the speed of light. From Eq.~(\ref{eq1}),  the corresponding group velocity of a photon with frequency $\nu$ becomes	
\begin{eqnarray}\label{eq2}
	v = \frac{\partial E}{\partial p} = c\sqrt{1 - \frac{ m_{\gamma}^2 c^4}{h_{p}^2 \nu^2}}\approx c\left(1-\frac{1}{2}A\nu^{-2}\right),
\end{eqnarray}
where $A \equiv \frac{m_{\gamma}^2 c^4}{h_{p}^2}$ and $h_p$ is the Planck constant.  The approximation is valid only for a very small photon mass, $m_\gamma \ll E/c^2 \approx 7\times10^{-42}\left(\frac{\nu}{\mathrm{GHz}}\right) \m{kg}$~\cite{Wei:2020wtf}.  Eq.~(\ref{eq2}) indicates that  if $m_{\gamma}\neq0$, the speed of light becomes frequency-dependent, with higher-frequency photons propagating faster.

We now consider an astrophysical source at redshift $z$ emitting radiation with high and low frequencies $\nu_{h}$ and $\nu_{l}$. If the high-frequency photons arrive at $z=0$, the low-frequency photons, traveling more slowly, arrive at time $-\Delta z$ (with $\Delta z\ll1$). The comoving distances traveled by the two components are
 $$x_{\nu_{h}}=\frac{c}{H_0}\int_{0}^{z} \left[ 1-\frac{1}{2}A\nu_{h}^{-2}(1+z')^{-2} \right] \frac{\mathrm{d}z'}{ \tilde{E} (z')}$$ and 
$$x_{\nu_{l}}= \frac{c}{H_0}\int_{-\Delta z}^{z} \left[ 1-\frac{1}{2}A\nu_{l}^{-2}(1+z')^{-2} \right] \frac{\mathrm{d}z'}{\tilde{E} (z')},$$
where $H_0$ is the Hubble constant, $ \tilde{E}(z)=H(z)/H_0$ is the dimensionless Hubble parameter. The factor $(1+z)^{-2}$ accounts for cosmic redshift of the photon frequency. 	Since both photon components originate from the same source, their comoving distances must be equal, i.e., $x_{\nu_{h}}=x_{\nu_{l}}$, which leads to 
\begin{eqnarray}
	\Delta z \simeq \frac{A}{2}(\nu_{l}^{-2} - \nu_{h}^{-2}) H_{\gamma}(z) , 
\end{eqnarray}
where
\begin{eqnarray}
	H_{\gamma}(z) \equiv \int_{0}^{z}\frac{(1+z')^{-2}}{ \tilde{E} (z')}\m{d}z'.
	\label{Hgamma}
\end{eqnarray}
The corresponding time delay induced by the photon mass is then 
\begin{eqnarray}
	\Delta t_{m_{\gamma}} = \frac{\Delta z}{H_0} \simeq  \frac{A}{2H_0}(\nu_{l}^{-2} - \nu_{h}^{-2})  H_{\gamma}(z).
\end{eqnarray}
The time delay due to $m_\gamma\neq 0$ can be interpreted as an effective DM~\cite{Shao:2017tuu, Wu:2016brq, Wei:2020wtf, Wang:2023fnn},  
\begin{eqnarray}
	\mathrm{DM_{\gamma}} = \Delta t_{m_{\gamma}} \frac{8\pi^2m_e \epsilon_0 c} {e^2}(\nu_{l}^{-2} - \nu_{h}^{-2})^{-1} = \frac{4\pi^2 m_e \epsilon_0 c^5}{h_{p}^2 e^2} \frac{H_\gamma (z)}{H_0} m_{\gamma}^2
	\label{dm_gam}.
\end{eqnarray}
Here $\epsilon_0$ represents the permittivity of vacuum,  $e$ is the electron charge, and $m_e$ is the mass of electron.   

By comparing  the observed DM with Eq.~(\ref{dm_gam}), we may constrain the photon rest mass. 
Since $H_{\gamma}(z)$ depends on the expansion history, adopting a cosmological model is essential. Motivated by recent DESI BAO results favoring dynamical dark energy, we consider three flat cosmological models, i.e.,  $\Lambda$CDM, $w$CDM and $w_{0}w_{a}$CDM. Their dimensionless Hubble parameters are	 
\begin{eqnarray}
	\tilde{E}(z) =
	\begin{cases}
		\sqrt{\Omega_{m0}(1+z)^3+\Omega_{r0}(1+z)^4+(1-\Omega_{m0}-\Omega_{r0})}& \Lambda\m{CDM}\\[8pt] \sqrt{\Omega_{m0}(1+z)^3+\Omega_{r0}(1+z)^4+(1-\Omega_{m0}-\Omega_{r0})(1+z)^{3(1+w)}}& w\m{CDM} \\[8pt]
		\sqrt{\Omega_{m0}(1+z)^3+\Omega_{r0}(1+z)^4+(1-\Omega_{m0}-\Omega_{r0})(1+z)^{3(1+w_{0}+w_{a})}e^{-3w_{a}\frac{z}{1+z}}}&  w_{0}w_{a}\m{CDM}.
	\end{cases}
	\label{model}
\end{eqnarray}
Here  $\Omega_{m0}$ and $\Omega_{r0}$ are the present matter and radiation density parameters, and $w$, $w_{0}$ and $w_{a}$  are constants characterizing dynamical dark energy. The radiation density is given by\begin{eqnarray}
	\Omega_{r0}=\Omega_{\gamma0}(1+0.2271N_{eff}),
\end{eqnarray}
where $ \Omega_{\gamma0}=\frac{3}{4h^2\cdot31500}\left(\frac{T_\m{CMB}}{2.7 \m{K}}\right)^4$ is the present photon energy density with $h\equiv H_0/100~\m{km} ~ \m{s}^{-1}~ \m{Mpc}^{-1}$ and $T_\m{CMB}=2.7255~\m{K}$~\cite{Chen:2018dbv}, and $N_{eff}=3.044$ represents the effective number of relativistic species.
To break degeneracies between cosmological parameters and the photon mass, we include complementary datasets:  SN Ia, CMB and BAO. 

\section{Observational Data}
\label{secdata}
\subsection{FRB}

For FRBs, the observed dispersion measure, $\mathrm{DM}_{\mathrm{obs}}$, contains contributions from several astrophysical sources as well as  a potential contribution from a non-zero photon mass: 
\begin{eqnarray}
	\m{DM}_\m{obs}(z) &=& \m{DM_{astro}} + \m{DM_{\gamma}} \nonumber\\
	&=&\m{DM}_\m{ISM}^\m{MW} + \m{DM}_\m{halo}^\m{MW}+ \m{DM}_\m{cosmic}(z) + \m{DM}_\m{host}/(1+z) + \m{DM_{\gamma}},
	\label{eq:comp}
\end{eqnarray}
where the superscript MW denotes the Milky Way, and the subscripts refer to contributions from the interstellar medium (ISM), the Milky Way halo, the cosmic baryon contribution (including the IGM and intervening halos~\cite{Connor:2024mjg}), the FRB host galaxy, and the photon-mass term.
	
Due to significant fluctuations in the distribution of baryons along the signal path, only the average cosmic dispersion measure $\mathrm{\langle DM_{cosmic}\rangle}$ can be computed reliably. 	Its theoretical expression is~\cite{Ioka:2003fr, Inoue:2003ga}:
\begin{eqnarray}
	\mathrm{\langle{DM_{cosmic}}\rangle}&=&\frac{3 c\Omega_{b0}  {H_0}f_\mathrm{{d}}}{8\pi G m_p}
	\int_0^z\frac{(1+z')\chi_e(z')}{\tilde{E}(z')}\mathrm{d}z'
	\label{aver_dmigm}
\end{eqnarray}
where $\Omega_{b0}$ is the present cosmic baryon density fraction parameter,   $m_p$ is the photon mass, and $f_\m{d}$  is the baryon mass fraction in the diffuse state.  Following~\cite{Liu:2025fdf, Lin:2023opv}, we treate  $f_\m{d}$ as a constant, 
despite the expectation of an increasing trend with redshift~\cite{prochaska2019probing, McQuinn:2013tmc}.  For redshifts $z<3$, hydrogen and helium are fully ionized, giving $\chi_{e}=7/8$~\cite{Becker:2010cu, Meiksin:2007rz},  which describes the number of free electrons per baryon. The Milky Way ISM contribution  $\m{DM_{ISM}^{MW}}$  can be subtracted from the observed dispersion measure using electron-density models such as NE2001 and YMW16~\cite{Cordes:2002wz, Yao:2017kcp}. In this work, we adopt  NE2001. 
For the Milky Way halo, Prochaska et al.~\cite{prochaska2019probing} estimates $\m{DM_{halo}^{MW}}$ to lie between 50 and 80 $ \m{pc~cm^{-3}}$. Therefore, we impose a Gaussian prior $\mathcal{N}(65,~15^2)~ (\m{pc~cm^{-3}})$ truncated to this interval, 
though some recent studies suggest that $\m{DM_{halo}^{MW}}$ may take a lower value \cite{DeepSynopticArrayTeam:2023suu, Liu:2026txt}.

Even after removing these Milky Way contributions, degeneracies remain among the cosmic, host, and photon-mass components. Following Macquart et al.~\cite{Macquart:2020lln}, we define $\m{DM_{ext}'}$ as $\m{DM}$ corrected for contributions from the ISM and the halo:
\begin{eqnarray}
	\m{DM_{ext}'} &=& \m{DM_{obs} - DM_{ISM}^{MW} -\m{DM_{halo}^{MW}} }\nonumber \\
	&=& \m{DM_{cosmic}} + \m{DM_{host}}/(1+z) + \m{DM_{\gamma}} \equiv \m{DM_{ext}} + \m{DM_{\gamma}},
\end{eqnarray}
The probability distribution function (PDF)  of  $\m{DM_{ext}}$,  including cosmic and host contributions, is  
\begin{eqnarray} \label{eq10}
	P(\mathrm{DM_{ext}}) &=& \int_0^{\m{(DM'_{ext}-DM_\gamma)(1+z)}}\dfrac{1}{\mathrm{\langle DM_{cosmic}\rangle}}P_{\mathrm{host}}(\mathrm{DM_{host}}) \nonumber \\
	&\times& P_{\mathrm{cosmic}}\left(\dfrac{\mathrm{DM'_{ext}-  DM_{host}/(1+z)-DM_\gamma}}{\mathrm{\langle DM_{cosmic}\rangle}}\right) \m{d DM_{host}}.
\end{eqnarray}
This expression originates from  Macquart et al.~\cite{Macquart:2020lln}, although they omitted the normalization  factor $\dfrac{1}{\mathrm{\langle DM_{cosmic}\rangle}}$~\cite{Zhang:2025wif,Zhuge:2025urk}.
Let  $\Delta \equiv \m{DM_{cosmic}}/\m{\langle DM_{cosmic}\rangle}$,  the cosmic-component PDF, $P_\m{cosmic}(\m{\Delta})$,  is found to satisfy
\begin{eqnarray}\label{13}
	P_{\mathrm{cosmic}}(\Delta)=A\Delta^{-\beta}\exp\left[-\dfrac{(\Delta^{-\alpha}-1)^2}{2\alpha^2\sigma_{\mathrm{cosmic}}^2}\right],~~~~~\Delta>0,
	\label{Delta}
\end{eqnarray}
where  $A$ is  the normalization constant, and $\alpha$, $\beta$ and $\sigma_\m{cosmic}$ are model parameters. This formula differs slightly from that proposed by Macquart et al.~\cite{Macquart:2020lln} (see Eq.~(\ref{macquart_P})) as we fix $C_0 = 1$.  In~\cite{Macquart:2020lln}, $\alpha$ and $\beta$ were both set to 3 and $\sigma_\m{cosmic}=F\times z^{-0.5}$, with $F$ being a free parameter. Based on the continuous $\m{DM_{cosmic}}$ catalog, Konietzka et al.~\cite{Konietzka:2025kdr} found that  the PDF of $\m{DM_{cosmic}}$ takes a  similar form to that of Eq.~(\ref{13}), with  parameters $\alpha\approx 1$ and $\beta \approx 3.3$, which  fits well with their simulation data.    In our analysis, we treat $\alpha$, $\beta$ and $\sigma_\m{cosmic}$ as free parameters and calibrate them with mock data. The Kolmogorov–Smirnov test indicates that	allowing all three parameters to vary provides a significantly better description of $\m{DM_{cosmic}}$ (see Appendix.~\ref{app} for details).

We model the host-galaxy DM, $P_\m{host}(\m{DM_{host}})$,  as a log-normal distribution
\begin{eqnarray}
	P_\mathrm{host}(\mathrm{DM_{host}})=\dfrac{1}{\sqrt{2\pi}\mathrm{DM_{host}\sigma_{host}}}\times \exp\mathrm{\left[-\dfrac{(\ln{DM_{host}}-\mu_{host})^2}{2\sigma_{host}^2}\right]}.
	\label{host}
\end{eqnarray}
The parameters $\mu_{\mathrm{host}}$ and $\sigma_{\mathrm{host}}$ are adopted from the IllustrisTNG simulations~\cite{Zhang:2020mgq} within $0.1 < z < 1.5$,\footnote{Note that \cite{Zhang:2020mgq} presents the mean and standard deviation for $\ln \left(\m{DM_{host}/(1+z)}\right)$, rather than for $\ln \m{DM_{host}}$.} and their values at each redshift are obtained via cubic-spline interpolation. It should be noted  that the IllustrisTNG simulation in~\cite{Zhang:2020mgq} is conducted within the framework of $\Lambda \m{CDM}$ cosmology.  Therefore,  potential incompatibilities may arise when estimating parameters in other cosmological models.  To investigate whether such incompatibilities affect the posterior constraints, we additionally treat $\mu_{\mathrm{host}}$ and $\sigma_{\mathrm{host}}$ as free parameters, similar to what was done in \cite{Macquart:2020lln, James:2021jbo}, and compare the corresponding results with those derived from the IllustrisTNG-based prescription.

From the FRB likelihood function,  $\ln \mathcal{L}_\m{FRB} = \ln P(\m{DM_{ext}})$, we can derive constraints on the photon mass and cosmological parameters from FRBs. Recently, a collection of  115 localized FRBs was compiled in~\cite{Gao:2025fcr}, with the highest redshift $z=1.354$. This large dataset significantly enhances our ability to constrain free parameters.
In this compilation, FRB190520B and FRB220831A are excluded due to their extremely high $\m{DM_{obs}}$, and FRB220319D is discarded because it has a negative $\m{DM_{ext}}$.  Additionally, FRB20221027A is excluded for its ambiguity in the host galaxy localization~\cite{Sharma:2024fsq}. Considering that $\m{DM_{halo}}$ is restricted to  $(50, 80)\,\m{pc~cm^{-3}}$, we impose the additional condition: $\m{DM_{obs} - DM_{ISM}^{MW} > 80}~\m{pc~cm^{-3}}$ as filters to ensure physical consistency. We also require $z>0.02$, because the parameters in the PDF start at $z=0.02$.  After applying these cuts, 104 FRBs remain for our analysis.	 	
 	

\subsection{SN Ia}
The SN Ia data utilized in our analysis come from  the  Pantheon+ compilation, which  consists of 1701 supernovae with redshifts in the range $z\in[0.00122,~2.26137]$~\cite{Scolnic:2021amr}.  To reduce the impact of peculiar velocities on very nearby events~\cite{Brout:2022vxf}, we discard all objects with redshift  $z\leq 0.01$,  leaving 1590 supernovae in our final sample. The distance module $\mu$ of SN Ia is defined as $\mu = m - M$, where  $m$ is the SN apparent magnitude and  $M$ is the absolute magnitude,  treated as a nuisance parameter that  is marginalized over in the analysis. The theoretical distance module for a SN Ia can be evaluated using the following formula:
\begin{eqnarray}
\mu_\m{th} =  25 + 5\log_{10}\left(\frac{D_\m{L}(z)}{\m{Mpc}}\right),
\end{eqnarray}
where $D_\m{L}$ is the luminosity distance, calculated as 
$D_\m{L} = \frac{c(1+z)}{H_0}\int_0^z\frac{dz'}{\tilde{E}(z')}$ in a spatially flat universe. The log-likelihood function for SN Ia can be expressed in terms of the observed distance modules and the corresponding model predictions:
\begin{eqnarray}
	\ln \mathcal{L}_\m{SN} 
	\propto 
	-\frac{1}{2}(\bm{\mu}_\m{obs}-\bm{\mu}_\m{th})^\m{T} ~\bm{\m{C}}_\m{SN}^{-1}~(\bm{\mu}_\m{obs}-\bm{\mu}_ \m{th})
\end{eqnarray}
with $\bm{\m{C}}_\m{SN}$   the covariance matrix of the observed distance moduli.

\subsection{CMB}
For CMB data, we use the Planck 2018 results~\cite{Planck:2018vyg}.  Instead of the CMB temperature  and polarization power spectra,  we consider 
three derived parameters:  the acoustic scale $l_\m{A}$, the shift parameter $R$ and $\Omega_{b0} h^2$ from the Planck 2018 data~\cite{Chen:2018dbv} for simplifying numerical calculation. 
The theoretical expression of $l_\m{A}$ is
\begin{eqnarray}
	l_\m{A}(z^*) = (1+z^*)\frac{\pi D_\m{A}(z^*)}{r_{s}(z^*)},
	\label{la}
\end{eqnarray}
where $z^*=1089.92$~\cite{Planck:2018vyg}  is the redshift at  the photon decoupling epoch,  $D_\m{A}=D_\m{L}/(1+z)^2$ is the angular diameter distance, and  $r_\m{s}$ is the comoving sound horizon, defined as 
\begin{eqnarray}
	r_{s}(z) = \frac{c}{H_0}\int_{0}^{1/(1+z)}\frac{da}{a^2 \tilde{E} (a)\sqrt{3 \left (1+\frac{3\Omega_{b0} h^2}{4\Omega_{\gamma0}h^2}a \right)}}. 
\end{eqnarray}
Here $a=1/(1+z)$ is the scale factor.  	The shift parameter $R$ takes the form 
\begin{eqnarray}
	R(z^*)=(1+z^*)\frac{D_\m{A}(z^*)\sqrt{\Omega_{m0}}H_0}{c}.
	\label{R}
\end{eqnarray}
 	
Introducing the  vector $\bm{A}_{\m{CMB}}=\{R,~l_\m{A},~\Omega_{b0}h^2 \}$, the Planck CMB observations give $\bm{A}_{\m{CMB},\m{obs}}=\{1.7502,~301.471,~0.02236 \}$.  Then we find that the expression of the log-likelihood function for the CMB data is  
\begin{eqnarray}
	\ln \mathcal{L}_\m{CMB}  \propto  -\frac{1}{2} (\bm{A}_{\m{CMB},\m{obs}}-\bm{A}_{\m{CMB}, \m{th}})^\m{T} ~\bm{\m{C}}_\m{CMB}^{-1}~(\bm{A}_{\m{CMB},\m{obs}}-\bm{A}_{\m{CMB}, \m{th}}),
\end{eqnarray}
with $\bm{\m{C}}_\m{CMB}$  the covariance matrix provided in Tab.~I in~\cite{Chen:2018dbv}.

\subsection{BAO}
The latest  BAO measurements, released by  the DESI~\cite{DESI:2025zgx}, are used in this paper. These data are obtained from more than 14 million galaxies and quasars drawn from the DESI DR2.  	The extracted BAO signals, which expressed as  $D_\m{M}(z)/r_{s}(z_\m{d})$, $D_\m{H}(z)/r_{s}(z_\m{d})$ and  $D_\m{V}(z)/r_{s}(z_\m{d})$, enable the measurement of key cosmological quantities, where $D_\m{M}(z)$, $D_\m{H}(z)$ and $D_\m{V}(z)$ are  the transverse comoving distance, the Hubble horizon, and the angle-averaged distance, respectively, and $r_{s} (z_\m{d})$ is the sound horizon at the drag epoch $z_\m{d}=1059.94$~\cite{Planck:2018vyg}. 
In a spatially flat universe, the transverse comoving distance, the Hubble horizon  and  the angle-averaged distance  respectively take the forms:
\begin{eqnarray}
	D_\m{M}(z) = \frac{c}{H_0}\int_{0}^{z}\frac{dz'}{ \tilde{E} (z')}, \quad
	D_\m{H}(z) = \frac{c}{H_0 \tilde{E}(z)}, \quad
	D_\m{V}(z)=\left[ \frac{cz}{H_0}\frac{D_\m{M}^2(z)}{\tilde{E}(z)}\right]^{1/3}.
\end{eqnarray}
Defining $\bm{A}_{\m{BAO}}=\{D_\m{M}(z)/r_{s}(z_\m{d}), D_\m{H}(z)/r_{s}(z_\m{d}), D_\m{V}(z)/r_{s}(z_\m{d})\}$,  we can express the  log-likelihood function of BAO datasets as 
\begin{eqnarray}
	\ln \mathcal{L}_\m{BAO} \propto -\frac{1}{2}(\bm{A}_{\m{BAO},\m{obs}}-\bm{A}_{\m{BAO}, \m{th}})^\m{T} ~\bm{\m{C}}_\m{BAO}^{-1}~(\bm{A}_{\m{BAO},\m{obs}}-\bm{A}_{\m{BAO}, \m{th}})
\end{eqnarray}
with the $\bm{A}_{\m{BAO},\m{obs}}$ and the covariance matrix $\bm{\m{C}}_\m{BAO}$ shown in Tab.~IV in~\cite{DESI:2025zgx}.



\begin{table}
	\centering
	\caption{Summary of the MCMC parameters, their prior distributions, and posterior constraints by adopting the $\Lambda \m{CDM}$ model, the $w\m{CDM}$ model and the $w_{0}w_{a}\m{CDM}$ model.}
	\begin{tabular}{ccccc}
		\hline
		Parameter & Prior & $\Lambda \m{CDM}$ & $w\m{CDM}$ & $w_{0}w_{a}\m{CDM}$ \\
		\hline
		$H_0(\m{km}/(\m{s}\cdot \m{Mpc}))$        &    (0, 150)                 &   $68.55\pm0.30$           &   $68.09\pm0.59$ & $67.79\pm0.59$\\
		$\Omega_{m0}$  &    (0, 1)                   &   
		$0.3026\pm0.0037$        &   $0.3059\pm0.0051$ & $0.3115\pm0.0057$\\
		$\Omega_bh^2$  &    (0, 0.05)                &   
		$0.02253\pm0.00012$      &   $0.02256\pm0.00013$  & $0.02245\pm0.00013$\\
		$f_\m{d}$    &    (0, 1)                   &   
		$0.945^{+0.051}_{-0.017}$    &   $0.943^{+0.050}_{-0.020}$ & $\geq0.936$  \\
		$\m{DM_{halo}^{MW}}(\m{pc~cm^{-3}})$ & $\mathcal{N}(65,~15^2)$ in (50, 80) & 
		$\leq55.6$ & $\leq55.6$ & $\leq55.6$\\
		$m_{\gamma}(\m{kg})$   &    (0, $2\times10^{-49}$) &   
		$\leq4.83\times10^{-51}$ &   $\leq4.71\times10^{-51}$ & $\leq4.86\times10^{-51}$ \\
		$w(w_0) $ & (-3, 1) &
		- & $-0.978\pm0.024$ & $-0.855\pm0.056$ \\
		$w_a $ & (-3, 2) &
		- & - & $-0.52^{+0.23}_{-0.21}$ \\
		\hline
	\end{tabular}
	\label{tab1}
\end{table}

\begin{figure}
	\centering
	\includegraphics[width=0.8\linewidth]{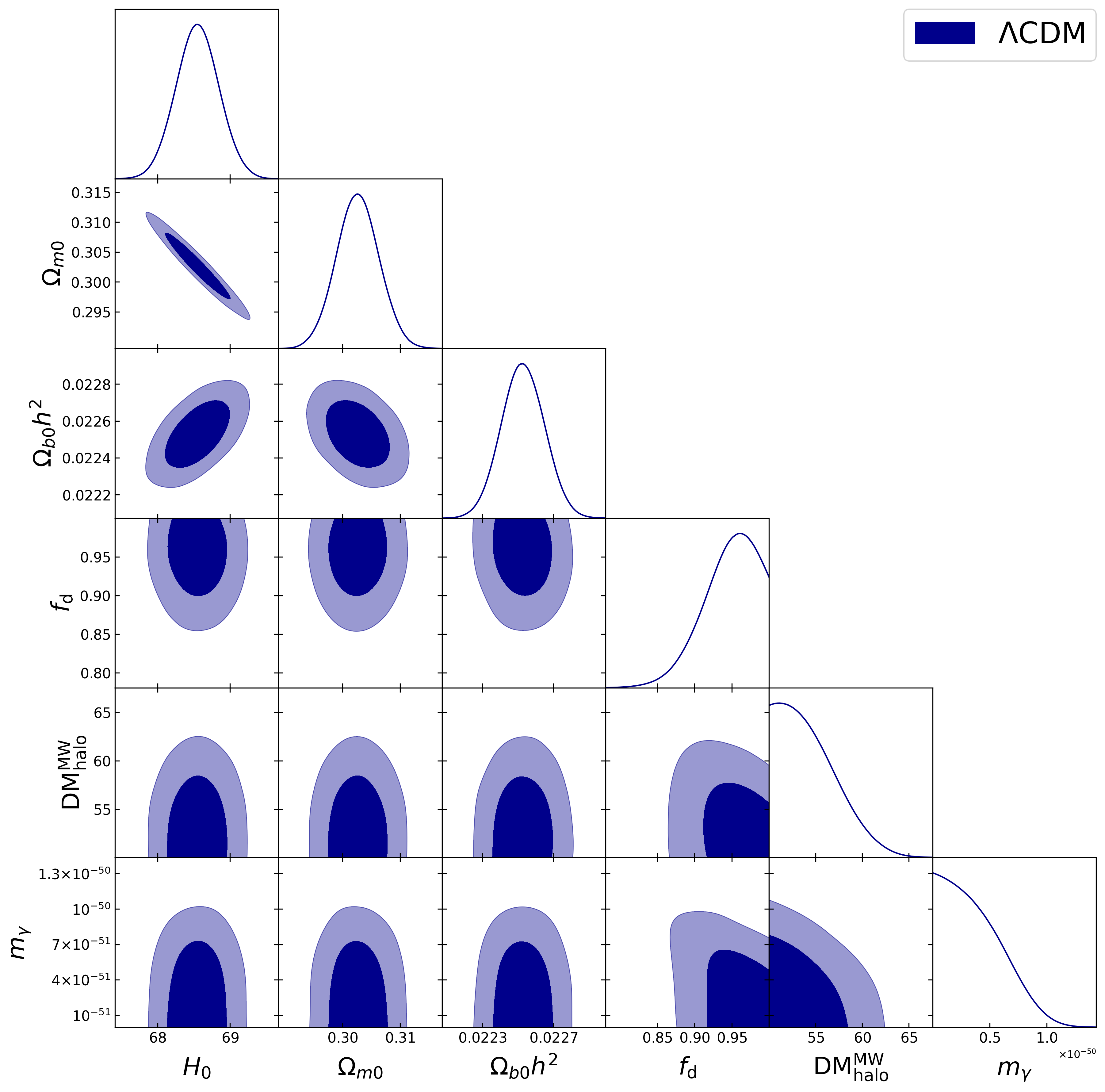}
	\caption{1D marginalized posterior distribution and 2D 1-2$\sigma$ contour regions for totally parameters 		 constrained by a combination of FRB, CMB, BAO and SN Ia in the $\Lambda \m{CDM}$ model.}
	\label{fig_L}
\end{figure}

\begin{figure}
	\centering
	\includegraphics[width=0.9\linewidth]{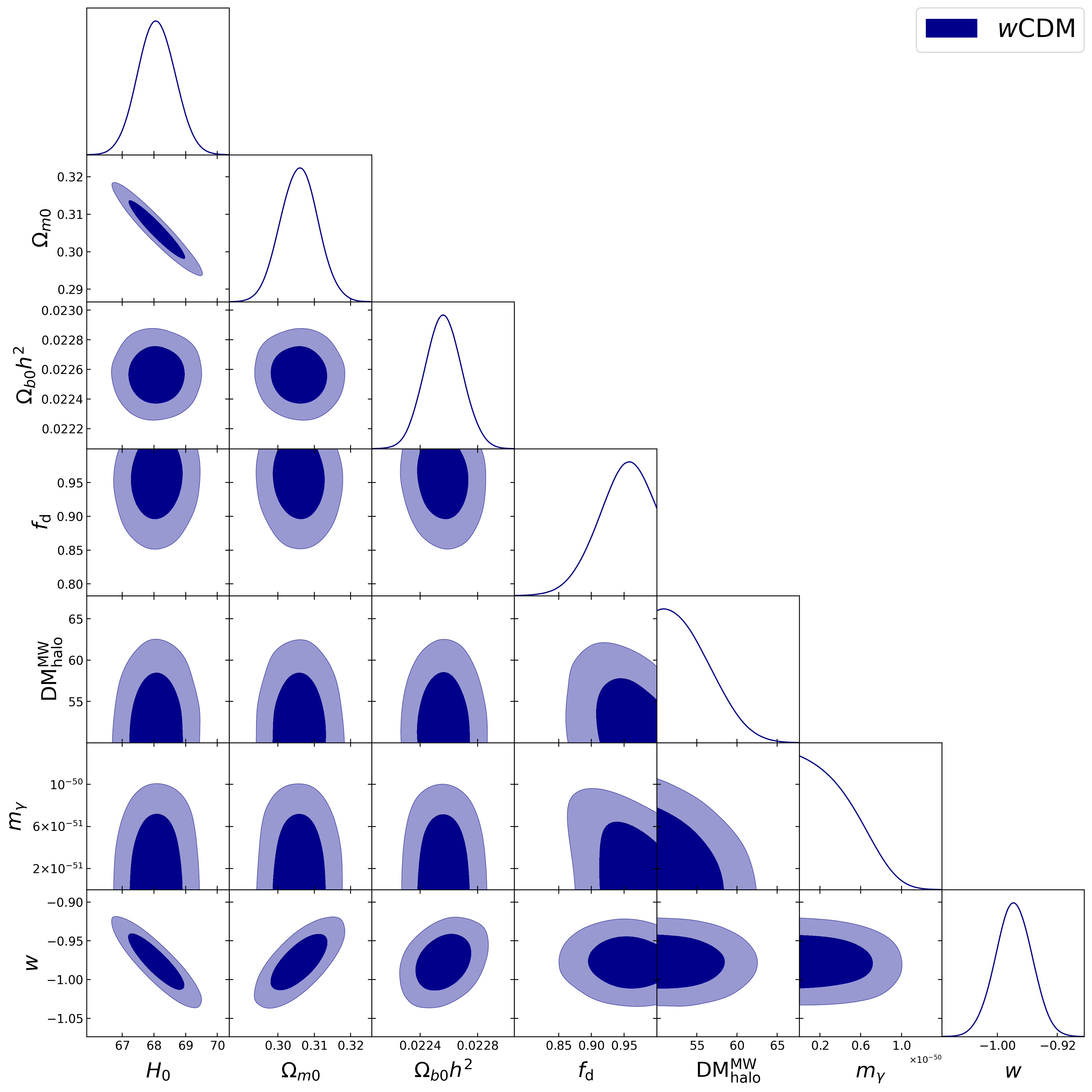}
	\caption{Constraints on photon mass and  free parameters  in the $w \m{CDM}$ model.}
	\label{fig_w}
\end{figure}

\begin{figure}
	\centering
	\includegraphics[width=1\linewidth]{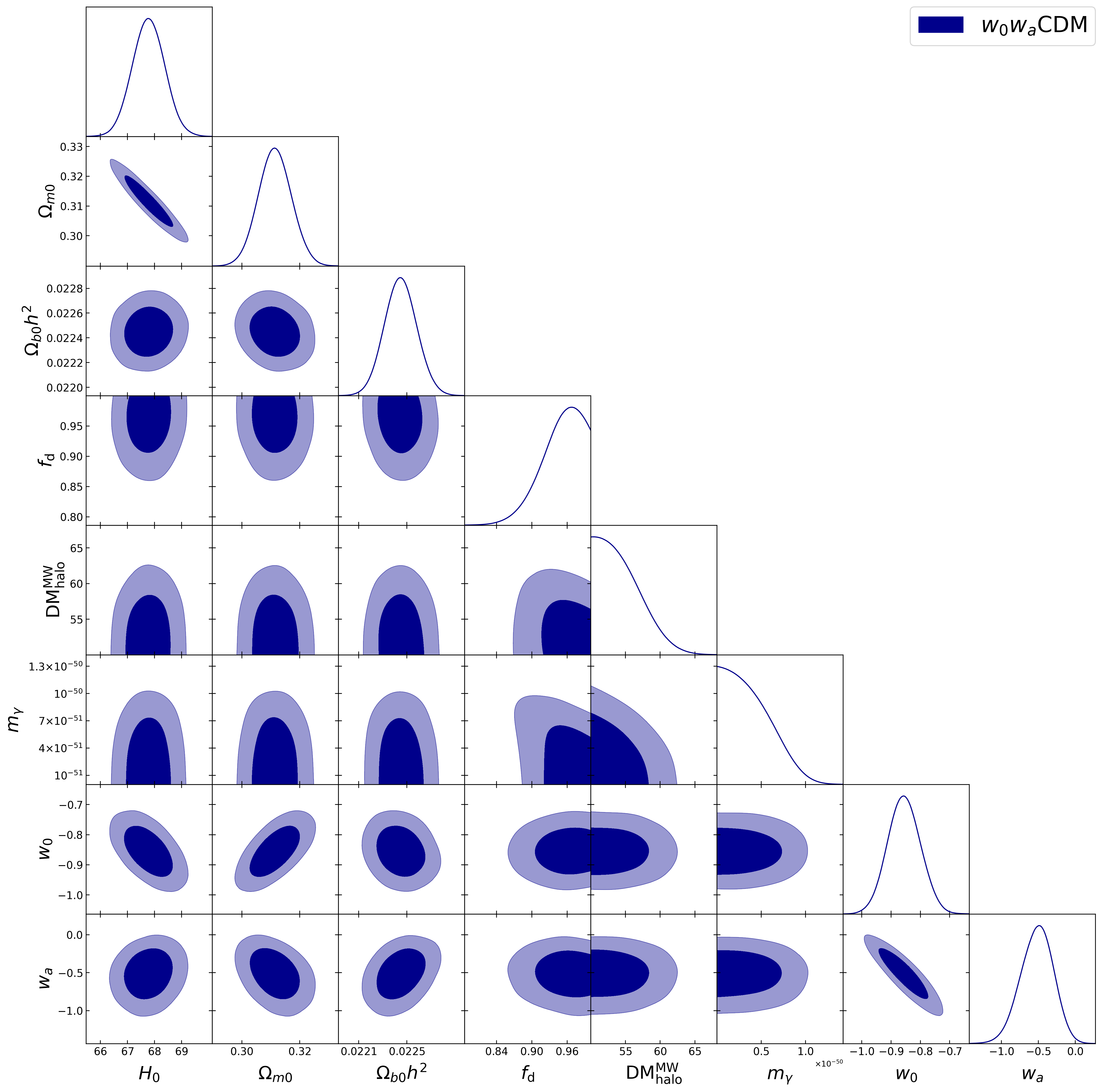}
	\caption{Constraints on  photon mass and free parameters  in the $w_{0}w_{a}\m{CDM}$ model.   		}
	\label{fig_cpl}
\end{figure}

\begin{table}
	\centering
	\caption{Summary of the MCMC parameters, including $\mu_{\m{host}}$ and $\sigma_{\m{host}}$, their prior distributions, and posterior constraints for the $\Lambda$CDM, $w$CDM, and $w_{0}w_{a}$CDM models.}
	\begin{tabular}{ccccc}
		\hline
		Parameter & Prior & $\Lambda \m{CDM}$ & $w\m{CDM}$ & $w_{0}w_{a}\m{CDM}$ \\
		\hline
		$H_0(\m{km}/(\m{s}\cdot \m{Mpc}))$        &    (0, 150)                 &   $68.56\pm0.29$           &   $68.13\pm0.58$ & $67.84\pm0.60$\\
		$\Omega_{m0}$  &    (0, 1)                   &   
		$0.3026\pm0.0037$        &   $0.3055\pm0.0051$ & $0.3111\pm0.0057$\\
		$\Omega_bh^2$  &    (0, 0.05)                &   
		$0.02253\pm0.00012$      &   $0.02256\pm0.00013$  & $0.02245\pm0.00013$\\
		$f_\m{d}$    &    (0, 1)                   &   
		$0.922^{+0.056}_{-0.035}$    &   $0.921^{+0.057}_{-0.036}$ & $0.927^{+0.060}_{-0.030}$  \\
		$\m{DM_{halo}^{MW}}(\m{pc~cm^{-3}})$ & $\mathcal{N}(65,~15^2)$ in (50, 80) & 
		$\leq53.8$ & $\leq53.8$ & $\leq53.9$\\
		$m_{\gamma}(\m{kg})$   &    (0, $2\times10^{-49}$) &   
		$\leq4.28\times10^{-51}$ &   $\leq4.26\times10^{-51}$ & $\leq4.22\times10^{-51}$ \\
		$w(w_0) $ & (-3, 1) &
		- & $-0.980\pm0.024$ & $-0.856\pm0.056$ \\
		$w_a $ & (-3, 2) &
		- & - & $-0.52^{+0.23}_{-0.21}$ \\
		$\mu_\m{host}$ & ($3.40$, $5.40$) &
		$4.33^{+0.22}_{-0.18}$ & $4.32^{+0.22}_{-0.18}$ & $4.33^{+0.22}_{-0.18}$ \\
		$\sigma_{\mathrm{host}}$ & (0.5, 1.5) &
		$1.10^{+0.11}_{-0.16}$ & $1.10^{+0.11}_{-0.16}$ & $1.10^{+0.11}_{-0.16}$ \\
		\hline
	\end{tabular}
	\label{tab_add}
\end{table}

When the FRB, SN Ia, CMB and BAO are considered together, the posterior distributions for free parameters can be acquired by maximizing the joint log-likelihood function:
\begin{eqnarray}
	\ln \mathcal{L}_\m{total} = \sum_i~\ln \mathcal{L}_i~,~i=\m{FRB,~SN~Ia,~CMB,~BAO}.
	\label{log_all}
\end{eqnarray} 
We employ emcee python package~\cite{Foreman-Mackey:2012any} for applying Markov Chain Monte Carlo (MCMC)  process, based on the joint log-likelihood function given in Eq.~(\ref{log_all}). The  priors for all parameters are listed in the second column of Tab.~\ref{tab1}.

\section{Results}
\label{result}

Within the framework of three cosmological models, we run 32 independent Markov Chain Monte Carlo (MCMC) chains for each model. The number of iterations ($\m{N_{steps}}$) for each chain is automatically controlled by the sampler. Convergence is assessed using the integrated autocorrelation time $\tau$ for all parameters,  with the stopping criterion:  (1)  $\m{N_{steps}} > 50\times \tau$ and  (2) $\left|\frac{\tau_\m{old}-\tau}{\tau}\right| < 0.01$, where $\tau_\m{old}$ is the value from the previous checkpoint.  The convergence test is performed every 1,000 iterations.   	
 	
In the context of the $\Lambda \m{CDM}$ model, the constraints on cosmological parameters and the photon mass are presented  in Fig.~\ref{fig_L} and summarized in Tab.~\ref{tab1}.  The combined  FRB, SN Ia, CMB and BAO yield tight constraints on the key cosmological parameters:  $H_0=68.55\pm 0.30~\mathrm{km/(s\cdot Mpc)}$, $\Omega_{m0}=0.3026\pm 0.0037$ and $\Omega_{b0}h^2=0.02253\pm 0.00012$ at the $1\sigma$ CL. These results are well consistent with previous determinations from Planck 2018, DESI, and ACT~\cite{Planck:2018vyg, DESI:2025zgx, ACT:2025fju}.  The baryon fraction in the diffuse state is constrained to  $f_\m{d}=0.945^{+0.051}_{-0.017}$ and the Milky Way halo dispersion measure satisfies $\m {DM_{halo}^{MW} }\leq 55.6$ at the $1\sigma$ CL. Most importantly,  we obtain an upper limit on the photon mass: $4.83\times 10^{-51}~\m{kg}$ at the $1\sigma$ CL. 

Fig.~\ref{fig_w} and Tab.~\ref{tab1} illustrate the constraints when the $w\m{CDM}$ model is adopted.  Compared  to $\Lambda \m{CDM}$, the constraints on  $H_0$, $\Omega_{m0}$,  $\Omega_{b0}h^2$,  $f_\m{d}$ and $\m {DM_{halo}^{MW} }$ change only slightly.   The equation of state $w$ for dark energy  is $w=-0.978\pm0.024$
which is marginally  above $-1$, though $w=-1$ remains allowed at the $1\sigma$ CL. The photon mass  constraint slightly improves to $m_{\gamma}<4.71\times 10^{-51}~\m{kg}$,  smaller than $4.83\times 10^{-51}~\m{kg}$ obtained in the $\Lambda$CDM model. Even with one additional free parameter relative to  $\Lambda \m{CDM}$,  the limit is essentially unchanged, indicating that the photon mass constraint is robust.	

We next consider the Chevallier-Polarski-Linder (CPL) parametrization of dark energy, with an evolving equation of state $ w(a) = w_0 + w_a (1-a)$. Results are shown in Fig.~\ref{fig_cpl} and Tab.~\ref{tab1}. 
The   constraints on  $H_0$, $\Omega_{m0}$,  $\Omega_{b0}h^2$,  $f_\m{d}$ and $\m {DM_{halo}^{MW} }$ remain consistent with those obtained in the $\Lambda$CDM and $w$CDM models.  We find that the allowed region for $w_a$ deviates from zero at more than $2\sigma$, indicating that a preference for dynamical dark energy.  In addition, the value of $w_0$ is noticeably larger than $-1$,  suggesting a lower present-day expansion rate of our universe than predicted  by the $\Lambda$CDM model. These trends are consistent with recent DESI DR2 findings~\cite{DESI:2025zgx, Feng:2025cwi, Smith:2025icl, DESI:2025fii}.  The upper limit on the photon mass is  $4.86\times 10^{-51}~\m{kg}$,
 differing only slightly from the values obtained in  $\Lambda$CDM and $w$CDM. This result might indicate that   the inferred photon mass exhibits little sensitivity to the assumed background cosmological model. We note, however, that this conclusion should be interpreted with some caution, as the current sample of localized FRBs is dominated by low-redshift sources, with only a small number located near $z \sim 1$.

To further assess the potential influence of   the assumption inherited from the $\Lambda\mathrm{CDM}$ -based IllustrisTNG simulation, we additionally treat $\mu_{\mathrm{host}}$ and $\sigma_{\mathrm{host}}$ as free parameters,  with their respective priors   being  $\mu_{\mathrm{host}} \in (3.40, 5.40)$ and $\sigma_{\mathrm{host}}\in (0.5, 1.5)$.  The results are summarized in Table.~\ref{tab_add}. The photon  mass is  constrained to  
 $m_{\gamma}\leq4.28\times10^{-51}~\m{kg}$, $m_{\gamma}\leq4.26\times10^{-51}~\m{kg}$ and $m_{\gamma}\leq4.22\times10^{-51}~\m{kg}$ 
in the contexts of $\Lambda \m{CDM}$, $w\m{CDM}$ and $w_{0}w_{a}\m{CDM}$ cosmologies, respectively.   Compared with results  in Tab.~\ref{tab1}, where $\mu_{\mathrm{host}}$ and $\sigma_{\mathrm{host}}$ are derived  from the IllustrisTNG simulation, we find that treating $\mu_{\mathrm{host}}$ and $\sigma_{\mathrm{host}}$ as free parameters leads to only minor changes in the inferred photon-mass bounds, in fact yielding slightly tighter constraints.
The variation in the baryon mass fraction $f_\m{d}$ resulting from two different treatments of $\mu_{\mathrm{host}}$ and $\sigma_{\mathrm{host}}$   is non-negligible, with a  smaller $f_\m{d}$ obtained when  $\mu_{\mathrm{host}}$ and $\sigma_{\mathrm{host}}$ are treated as free parameters. By contrast, the remaining  cosmological  parameters, including $H_0$, $\Omega_{m0}$, $\Omega_b h^2$, $\m{DM^{MW}_{halo}}$, $w(w_0)$ and $w_a$, remain in good agreement with 
those derived using the IllustrisTNG-based priors.


\section{Conclusions and discussions}
\label{seccon}
The DMs of cosmological and well-localized FRBs have  become powerful tools for estimating cosmological parameters. In this paper, we first improved  the DM distribution from extragalactic gas proposed in~\cite{Macquart:2020lln}   by restoring a missing normalization factor and by allowing the parameters $\alpha$ and $\beta$ in the probability distribution function for the DM of cosmic origin  to vary with redshift. We then utilized this improved distribution to test  whether the photon has a non-zero rest mass, since any such mass would contribute an additional effective DM term.  Using 104 localized FRBs, combined with SN Ia, CMB, and BAO data,
we obtain  the following $1\sigma$ upper limits on the photon rest mass: $m_{\gamma}\leq4.83\times10^{-51}~\m{kg}$ for $\Lambda \m{CDM}$ model, $m_{\gamma}\leq4.71\times10^{-51}~\m{kg}$ for the $w\m{CDM}$ model and $m_{\gamma}\leq4.86\times10^{-51}~\m{kg}$ for the $w_{0}w_{a}\m{CDM}$ model.  These values differ only minimally among the three cosmological models.  However, we refrain from concluding  that the inferred photon mass is entirely insensitive to the assumed cosmological background, since the current FRB sample is dominated by low-redshift sources, with only a few localized near $z\sim1$.   We further assess the impact of the $\Lambda$CDM assumption adopted in the IllustrisTNG simulation  by treating  $\mu_{\mathrm{host}}$ and $\sigma_{\mathrm{host}}$ as free parameters.  In this case, the photon mass is constrained to $m_{\gamma}\leq4.28\times10^{-51}~\m{kg}$, $m_{\gamma}\leq4.26\times10^{-51}~\m{kg}$ and $m_{\gamma}\leq4.22\times10^{-51}~\m{kg}$ within the $\Lambda \m{CDM}$, $w\m{CDM}$ and $w_{0}w_{a}\m{CDM}$ cosmologies, respectively.  These constraints are slightly tighter than those obtained when  $\mu_{\mathrm{host}}$ and $\sigma_{\mathrm{host}}$ are fixed to  the IllustrisTNG-based values.   Overall, our analysis  provides strong  empirical support for the massless nature of the photon. 

\begin{figure}[htbp]
	\centering
	\includegraphics[width=0.8\textwidth]{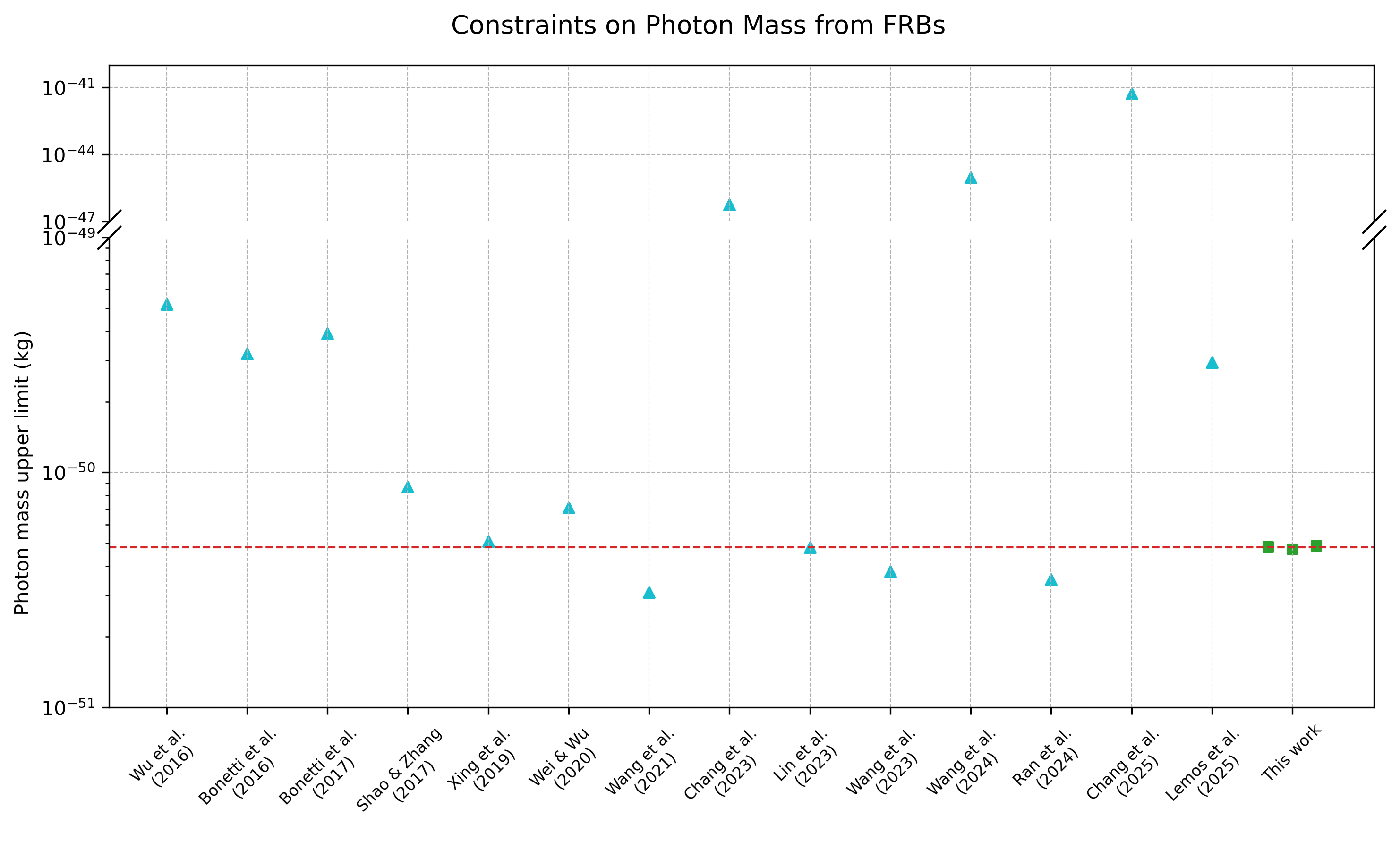}
	\caption{The upper limits on the photon rest mass from FRBs, derived from both  previous works and the current study, are presented. Cyan points represent constraints  obtained from previous studies, while the three green ones correspond to our results within the framework of three different cosmological models. The red dashed line indicates the average upper limit on $m_\gamma$ across the three cosmological models considered in this work.}
	\label{all}
\end{figure}

Fig.~\ref{all}  compares our photon-mass constraints with previous FRB-based limits.  Each cyan point represents an upper limit obtained in earlier studies~\cite{Wu:2016brq, Bonetti:2016cpo, Bonetti:2017pym, Shao:2017tuu, Xing:2019geq, Wei:2020wtf, Wang:2021nrl, Chang:2022qct, Lin:2023jaq, Wang:2023fnn, Lemos:2025qyh, Chang:2024hnn, Ran:2024avn, Wang:2024rgu}, while the three green points correspond  to the upper limits obtained here.  The red dashed line shows the average upper limit on $m_\gamma$ across three cosmological models we considered. It is apparent  that our results are  weaker than  previous values such as  $m_{\gamma}\leq3.1\times10^{-51}~\m{kg}$ (from 129 FRBs,  most of which lack  redshift measurements~\cite{Wang:2021nrl}),   $m_{\gamma}\leq3.5\times10^{-51}~\m{kg}$~\cite{Ran:2024avn} (from 32 well-localized FRBs using  a cosmological-independent method to determine the luminosity distance), and $m_{\gamma}\leq3.8\times10^{-51}~\m{kg}$~\cite{Wang:2023fnn} (from 32 well-localized FRBs with the cosmological parameters determined from the CMB, SNIa and BAO). 

However, these earlier studies~\cite{Wang:2021nrl, Ran:2024avn,Wang:2023fnn}   employed the $\m{DM_{ext}}$ distribution proposed by Macquart et al.~\cite{Macquart:2020lln}, which omits an essential normalization factor $1/\langle \m{DM_{IGM}}\rangle$. As shown in recent analyses~\cite{Zhang:2025wif, Zhuge:2025urk}, this omission significantly biases  constraints on cosmological parameters. The  tight limits obtained in~\cite{Wang:2021nrl, Ran:2024avn,Wang:2023fnn} , even with smaller datasets, further illustrate the impact of the missing normalization. Our analysis incorporates the corrected distribution and therefore provides more reliable constraints.

\begin{acknowledgments}

We appreciate very much the insightful comments and helpful suggestions by the anonymous referee. 	The authors Yuchen Zhang and Yang Liu make equal contributions to this work. This work was supported by the National Natural Science Foundation of China (Grants  No.~12275080 and No.~12075084), the Major basic research project of Hunan Province (Grant No.~2024JC0001), and the Innovative Research Group of Hunan Province (Grant No.~2024JJ1006).
	
\end{acknowledgments}

\bibliography{ref}

\appendix
\section{Modified probability distribution function of $\Delta$}
\label{app}
Since the distribution of free electrons in the IGM is inhomogeneous, the actual value of $\m{DM_{cosmic}}$ fluctuates around its average value $\langle\m{DM_{cosmic}}\rangle$.  To account for this variation,  Macquart et al.~\citep{Macquart:2020lln} introduced  $\Delta=\m{DM_{cosmic}}/\langle\m{DM_{cosmic}}\rangle$ and proposed a probability distribution function ({\it Macquart  distribution}) for  $\Delta$
\begin{eqnarray}
	P_{\mathrm{cosmic}}(\Delta)= A\Delta^{-\beta}\exp\left[-\dfrac{(\Delta^{-\alpha}-C_0)^2}{2\alpha^2\sigma_\m{cosmic}^2}\right].
	\label{macquart_P}
\end{eqnarray}
Here both $\alpha$ and $\beta $ are fixed to 3, $A$ is the normalization constant, $C_0$ is a parameter determined by the requirement that  the mean value of $\Delta$ equals 1, and $\sigma_\m{cosmic}$ is assumed to follow $\sigma_\m{cosmic}=Fz^{-1/2}$ with $F$ being  treated as a free parameter. 
Later, Zhang et al.~\cite{Zhang:2020xoc} treated $\sigma_\m{cosmic}$ in Eq.~(\ref{macquart_P}) as a free parameter and employed the IllustrisTNG cosmic simulation to determine the optimal values for different redshift points. Their results have since been widely adopted in subsequent studies.	
Recently, Konietzka et al.~\cite{Konietzka:2025kdr} proposed a new method for measuring $\m{DM_{cosmic}}$ in the IllustirsTNG cosmic simulation and found Zhang et al.~\cite{Zhang:2020xoc} had misestimated the standard deviation and higher moments of the $\m{DM_{cosmic}}$ distribution by over 50\%. Based on the continuous $\m{DM_{cosmic}}$ catalog acquired through their new method, Konietzka et al.~\cite{Konietzka:2025kdr} found that  the PDF of $\m{DM_{cosmic}}$ takes a form  similar to that of Eq.~(\ref{macquart_P}), with  parameters $\alpha\approx 1$ and $\beta \approx 3.3$ ({\it Konietzka  distribution}), which  fits well with their simulation data.   However, since the Konietzka  distribution~\cite{Konietzka:2025kdr} does not explicitly include the average $\langle\m{DM_{cosmic}}\rangle$, which encodes cosmological information, it is therefore incapable of using in FRB cosmology.

\begin{figure}[htbp]
	\centering
	\includegraphics[width=0.8\textwidth]{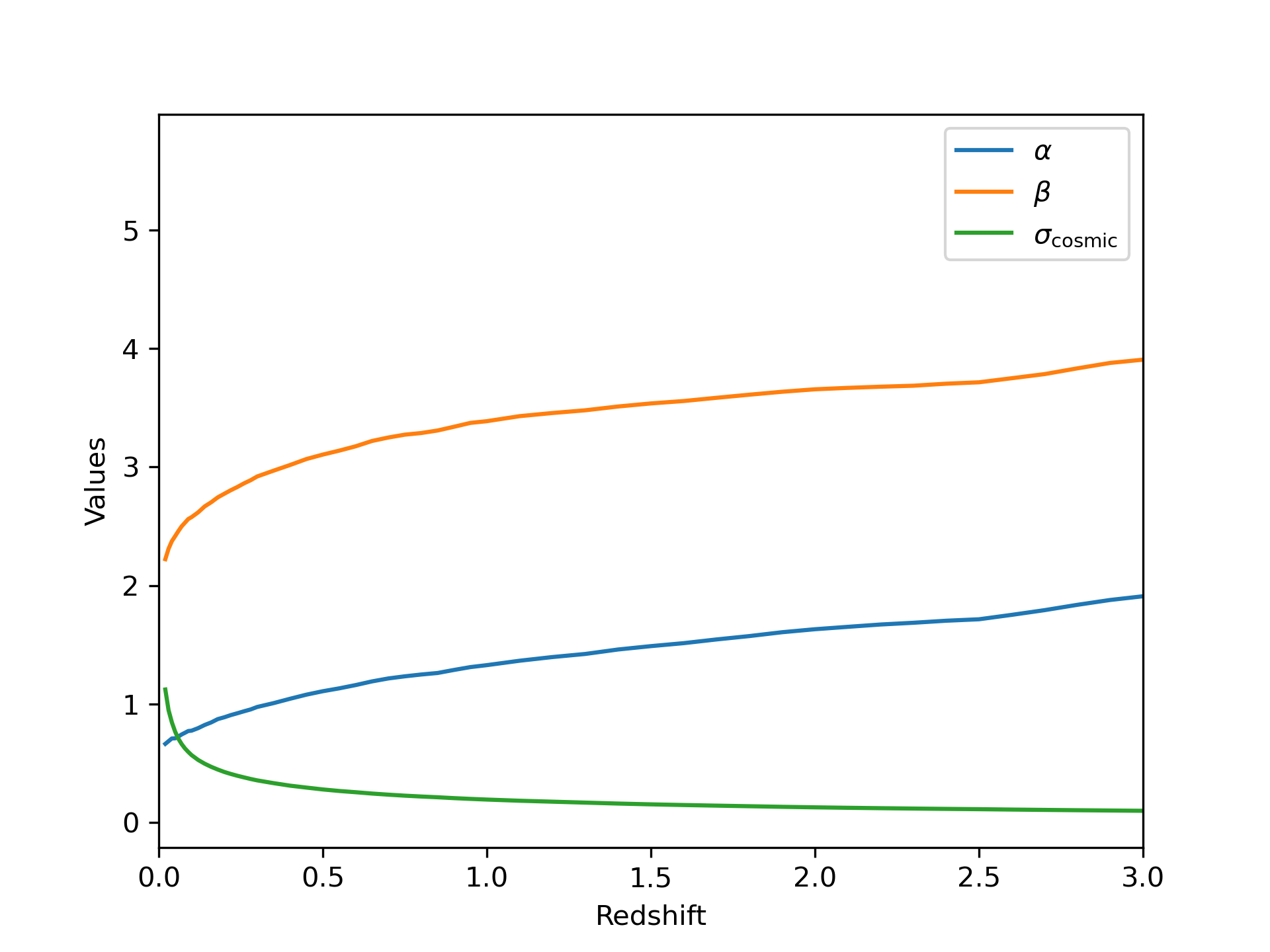}
	\caption{Variations of $\alpha$, $\beta$ and $\sigma_\m{cosmic}$ with redshift $z$.}
	\label{para}
\end{figure}

To address this problem, we assume that $P_\m{cosmic}(\Delta)$ takes a form similar as Eq.~(\ref{macquart_P}), but $C_0$ is fixed to 1, 
\begin{eqnarray}\label{A2}
	P_{\mathrm{cosmic}}(\Delta)= A\Delta^{-\beta}\exp\left[-\dfrac{(\Delta^{-\alpha}-1)^2}{2\alpha^2\sigma_\m{cosmic}^2}\right].
\end{eqnarray}
Since Konietzka et al.~\cite{Konietzka:2025kdr} found that $\alpha$ and $\beta$ taking  values  different from $\alpha=\beta=3$ provide better fit, we treat $\alpha$ and  $\beta$ as free parameters and determine their values using the continuous $\m{DM_{cosmic}}$ catalog.  In Eq.~(\ref{A2}), $\sigma_\m{cosmic}$ is also assumed to a free parameter,  determined along with  $\alpha$ and  $\beta$. To derive the values of $\alpha$,  $\beta$ and $\sigma_\m{cosmic}$, we first calculate the $\langle \m{DM_{cosmic}}\rangle$ from Eq.~(\ref{aver_dmigm}) for each redshift points, adopting $H_{0}=67.74~\m{km/s\cdot Mpc^{-1}}$, $\Omega_{m0}=0.3089$, $\Omega_{b0}=0.0486$ and $f_{eb}=f_{d}\cdot\chi_{e}=0.83$, which  are consistent with the assumptions made in generating the continuous catalog in~\cite{Konietzka:2025kdr}. Next, we employ a MCMC process to determine $\alpha$, $\beta$ and $\sigma_\m{cosmic}$ by fitting $P(\m{DM_{cosmic}})$ to the simulated $\m{DM_{cosmic}}$ distribution. Here we emphasize that the PDF for  $\m{DM_{cosmic}}$ is given by $P(\m{DM_{cosmic}})=\frac{1}{\langle\m{DM_{cosmic}}\rangle}P_\m{cosmic}(\Delta)$ as investigated in~\cite{Zhang:2025wif, Zhuge:2025urk}. The obtained parameters values for each redshift point are listed in Table.~\ref{tab_para}.
In Fig.~\ref{para},  we plot the variation of $\alpha$, $\beta$, and $\sigma_\m{cosmic}$ with redshift.  These parameters exhibit more significant variations at low redshift, while their evolution becomes flatter at higher redshifts. Therefore, we adopt finer redshift intervals at low $z$ and coarser intervals at high $z$.

\begin{figure}[htbp]
	\centering
	\includegraphics[width=0.8\textwidth]{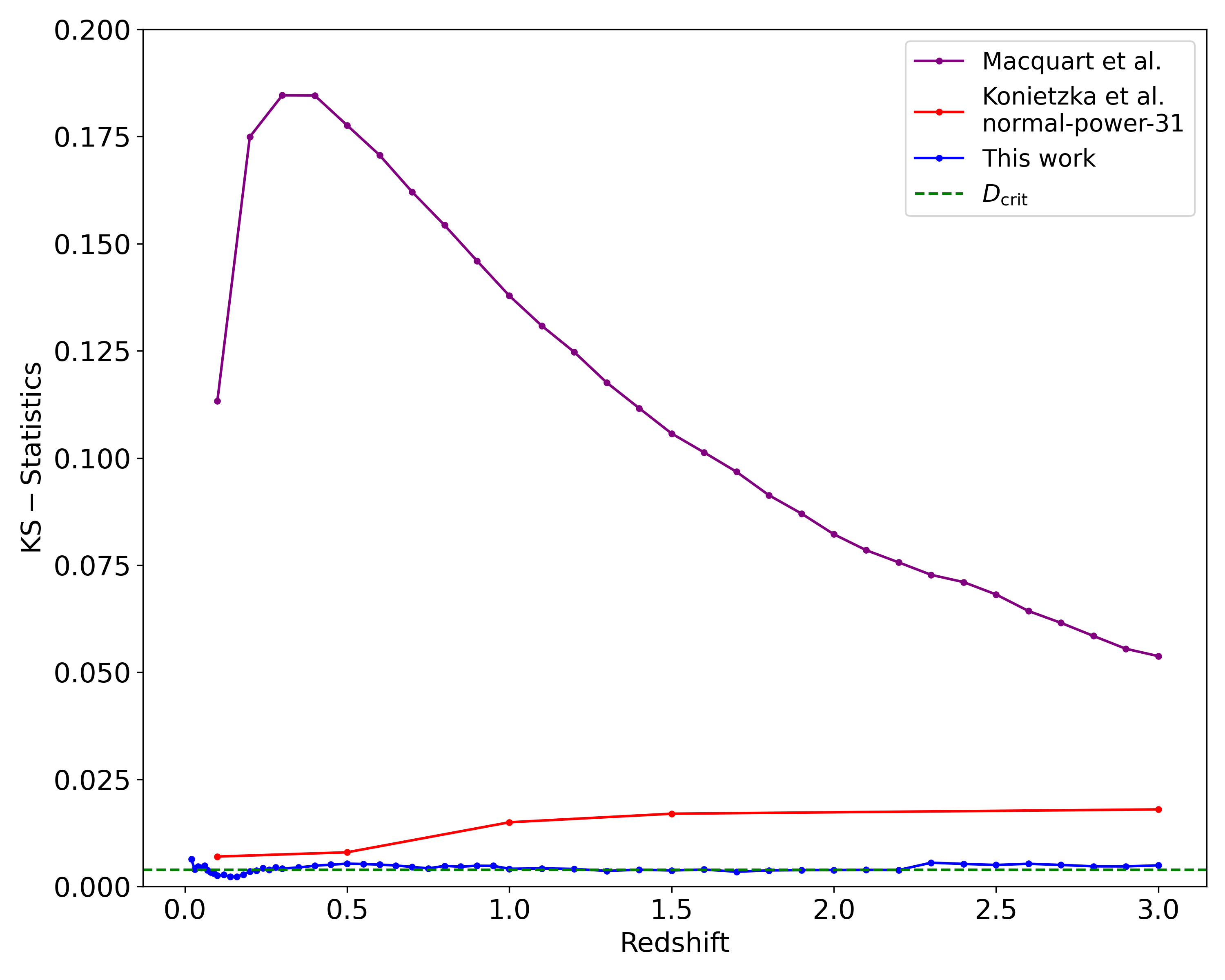}
	\caption{The KS statistical values versus redshift for three different probability functions of $\m{DM_{cosmic}}$.}
	\label{KS}
\end{figure}

To evaluate the validity of $P(\m{DM_{cosmic}})$ proposed here, we apply the  Kolmogorov–Smirnov (KS) test through  using the \texttt{SciPy} python package~\cite{2020SciPy-NMeth}.
The KS test is defined as, 
\begin{eqnarray}
	D_n(p) = \displaystyle \sup_{x} |F_p(x) - F_n(x)|,
\end{eqnarray}
where $F_p(x)$ and $F_n(x)$ are cumulative distribution functions (CDFs) corresponding to $P(\m{DM_{cosmic}})$ and the simulated $\m{DM_\m{cosmic}}$ distribution, respectively. The KS statistic $D_n(p)$ quantifies how far the samples deviate from the expected PDF. The KS test has a null hypothesis that the samples are drawn from $P(\m{DM_{cosmic}})$. We reject the null hypothesis if the difference  $D_n(p)$ exceeds a critical value $D_\m{crit}$, which can be calculated as $D_\m{crit}\approx \frac{c(\tilde{\alpha})}{\sqrt{n}}$, where $\tilde{\alpha}$ is the significance level that represents the probability of wrongly rejecting a true null hypothesis, and $n$ is the number of samples. For a specific redshift point, the continuous catalog has 120000 $\m{DM_{cosmic}}$ samples, and we adopt $\tilde{\alpha} = 0.05$ by convention.  Thus we obtain $c(\tilde{\alpha})=1.36$ (from {\it Real Statistics Using Excel}~\footnote{https://real-statistics.com/statistics-tables/kolmogorov-smirnov-table/}), leading to $D_\m{crit}\approx 0.0039 $. After applying the KS test to $P(\m{DM_{cosmic}})$ with the parameters listed in Table.~\ref{tab_para}, we plot the KS statistics for each redshift point in Fig.~\ref{KS}. The mean value of  the KS statistic is 0.0042, which slightly exceeds the $D_\m{crit}$.  We also include the KS statistics of the Macquart distribution~\cite{Macquart:2020lln} and the Konietzka  distribution~\cite{Konietzka:2025kdr} in Fig.~\ref{KS}. It is evident  that  the PDF in~\cite{Konietzka:2025kdr} performs  better than the one  in~\cite{Macquart:2020lln}, which is consistent with the finding  reported in~\cite{Konietzka:2025kdr}.  However, the KS statistics for both distributions clearly exceed  $D_\m{crit}$, indicating that  they are inadequate for describing the distribution of $\m{DM_{cosmic}}$.    Furthermore, we find that the parameter $F$ in Eq.~(\ref{macquart_P}) varies from $1.15$ to $0.16$ as the redshift changes from $0.1$ to $3.0$, suggesting  that $F$ should not be treated as a constant. This implies that  the relation $\sigma_\m{cosmic}=Fz^{-1/2}$ may be unsuitable.

\begin{table}[htbp]
	\centering
	\caption{Optimal values of parameters in $P_\mathrm{cosmic}(\Delta)$ based on the continuous catalog.}
	\begin{minipage}{0.48\textwidth}
		\centering
		\begin{tabular}{ccccc}
			\hline
			$z$ & $\alpha$ & $\beta$ & $\sigma_\mathrm{cosmic}$ & $A$ \\
			\hline
			0.02 & 0.6647 & 2.2223 & 1.1232 & 0.3612\\
			0.03 & 0.6876 & 2.3111 & 0.9473 & 0.4235\\
			0.04 & 0.7104 & 2.3746 & 0.8462 & 0.4726\\
			0.05 & 0.7113 & 2.4166 & 0.7675 & 0.5174\\
			0.06 & 0.7270 & 2.4599 & 0.7073 & 0.5610\\
			0.07 & 0.7444 & 2.5001 & 0.6623 & 0.5992\\
			0.08 & 0.7584 & 2.5305 & 0.6259 & 0.6344\\
			0.09 & 0.7731 & 2.5611 & 0.5966 & 0.6658\\
			0.10 & 0.7759 & 2.5770 & 0.5701 & 0.6963\\
			0.12 & 0.7973 & 2.6172 & 0.5292 & 0.7509\\
			0.14 & 0.8240 & 2.6674 & 0.4974 & 0.7997\\
			0.16 & 0.8458 & 2.7028 & 0.4708 & 0.8458\\
			0.18 & 0.8737 & 2.7439 & 0.4481 & 0.8902\\
			0.20 & 0.8892 & 2.7745 & 0.4275 & 0.9331\\
			0.22 & 0.9077 & 2.8045 & 0.4107 & 0.9719\\
			0.24 & 0.9232 & 2.8317 & 0.3954 & 1.0099\\
			0.26 & 0.9396 & 2.8620 & 0.3818 & 1.0459\\
			0.28 & 0.9550 & 2.8882 & 0.3686 & 1.0837\\
			0.30 & 0.9759 & 2.9199 & 0.3568 & 1.1202\\
			0.35 & 1.0083 & 2.9690 & 0.3334 & 1.1997\\
			0.40 & 1.0449 & 3.0165 & 0.3123 & 1.2823\\
			0.45 & 1.0797 & 3.0669 & 0.2960 & 1.3534\\
			0.50 & 1.1087 & 3.1045 & 0.2803 & 1.4298\\
			0.55 & 1.1329 & 3.1378 & 0.2674 & 1.4991\\
			0.60 & 1.1605 & 3.1743 & 0.2569 & 1.5611\\
			0.65 & 1.1916 & 3.2193 & 0.2457 & 1.6325\\
			0.70 & 1.2165 & 3.2487 & 0.2361 & 1.6991\\
			\hline
		\end{tabular}
	\end{minipage}%
	\hfill
	\begin{minipage}{0.48\textwidth}
		\centering
		\begin{tabular}{ccccc}
			\hline
			$z$ & $\alpha$ & $\beta$ & $\sigma_\mathrm{cosmic}$ & $A$ \\
			\hline
			0.75 & 1.2344 & 3.2725 & 0.2279 & 1.7608\\
			0.80 & 1.2497 & 3.2864 & 0.2206 & 1.8185\\
			0.85 & 1.2625 & 3.3078 & 0.2143 & 1.8716\\
			0.90 & 1.2882 & 3.3393 & 0.2069 & 1.9388\\
			0.95 & 1.3122 & 3.3718 & 0.2006 & 1.9999\\
			1.00 & 1.3286 & 3.3863 & 0.1950 & 2.0579\\
			1.10 & 1.3659 & 3.4286 & 0.1851 & 2.1684\\
			1.20 & 1.3969 & 3.4552 & 0.1771 & 2.2661\\
			1.30 & 1.4230 & 3.4781 & 0.1692 & 2.3720\\
			1.40 & 1.4607 & 3.5101 & 0.1613 & 2.4883\\
			1.50 & 1.4890 & 3.5360 & 0.1545 & 2.5970\\
			1.60 & 1.5141 & 3.5563 & 0.1489 & 2.6954\\
			1.70 & 1.5454 & 3.5835 & 0.1437 & 2.7913\\
			1.80 & 1.5738 & 3.6096 & 0.1387 & 2.8922\\
			1.90 & 1.6063 & 3.6345 & 0.1338 & 2.9988\\
			2.00 & 1.6316 & 3.6552 & 0.1294 & 3.0988\\
			2.10 & 1.6517 & 3.6672 & 0.1256 & 3.1941\\
			2.20 & 1.6714 & 3.6770 & 0.1222 & 3.2824\\
			2.30 & 1.6861 & 3.6853 & 0.1191 & 3.3671\\
			2.40 & 1.7033 & 3.7024 & 0.1163 & 3.4484\\
			2.50 & 1.7155 & 3.7142 & 0.1138 & 3.5240\\
			2.60 & 1.7529 & 3.7486 & 0.1106 & 3.6232\\
			2.70 & 1.7924 & 3.7834 & 0.1077 & 3.7209\\
			2.80 & 1.8375 & 3.8319 & 0.1050 & 3.8168\\
			2.90 & 1.8784 & 3.8776 & 0.1026 & 3.9077\\
			3.00 & 1.9096 & 3.9050 & 0.1003 & 3.9963\\
			\hline
		\end{tabular}
	\end{minipage}
	\label{tab_para}
\end{table}

%
%

\end{document}